\documentclass[a4paper,11pt]{article}
\pdfoutput=1

\usepackage{jcappub}
\usepackage[T1]{fontenc}
\usepackage{placeins}

\notoc

\title{Coherent multipath wave response on Reissner-Nordstr\"{o}m analogue surface}

\author[a]{Peng-Yu Chen,}
\author[b]{Sen Guo,}
\author[a]{Qing-Quan Jiang,}
\author[c]{Yu Liang,}
\author[d,e]{\mbox{and Kai Lin}}

\affiliation[a]{School of Physics and Astronomy, China West Normal University,\\
Nanchong 637000, People's Republic of China}
\affiliation[b]{College of Physics and Optoelectronic Engineering,\\
Chongqing Normal University, Chongqing 401331,\\
People's Republic of China}
\affiliation[c]{School of Big Data and Artificial Intelligence,\\
Fuyang University of Technology, Fuyang 236000, People's Republic of China}
\affiliation[d]{Universidade Federal de Campina Grande,\\
Campina Grande, PB, Brasil}
\affiliation[e]{Instituto de F\'isica, Universidade de S\~ao Paulo,\\
S\~ao Paulo, Brasil}

\emailAdd{sguophys@126.com}

\abstract{%
To uncover how the intrinsic metric of a relativistic compact object governs macroscopic wave phenomena, we establish a theoretical framework mapping the charge dependent spatial geometry of a Reissner-Nordstr\"{o}m (RN) black hole onto the coherent response of an analogue curved surface. By solving an exact spatial geodesic boundary value problem on an isometrically embedded equatorial slice, we extract the discrete multi-loop path-length spectrum and convert this geometric backbone into a physical wave field via a finite-path surface Huygens-Fresnel construction. We analytically compute the arbitrary order winding trajectories alongside their high winding accumulation limits, demonstrating that the analogue charge acts as a precise physical dial that reconfigures the event horizon throat and fundamentally reorganizes the discrete path sequence. Furthermore, we find that this underlying geometric deformation uniquely dictates the macroscopic interference, revealing that steady state spatial fringes, spectral resonance combs, and transient temporal echo ladders are intrinsically unified physical projections of a single charge controlled path spectrum. This systematic parameter to response methodology establishes a rigorous theoretical bridge between strong field gravitational lensing and tabletop transformation optics, providing a highly tunable blueprint for future multi domain analogue gravity experiments.
}


\hypersetup{
  pdftitle={Coherent multipath wave response on Reissner-Nordstrom analogue surface},
  pdfauthor={Peng-Yu Chen; Sen Guo; Qing-Quan Jiang; Yu Liang; Kai Lin},
  pdfkeywords={analogue optics; parameter tuning; optical echoes; spatial interference}
}

\begin{document}
\maketitle
\flushbottom

\section{Introduction}
\par
General relativity elegantly interprets gravity as the geometric manifestation of curved spacetime, where the underlying metric dictates the structure of light cones and governs the propagation of fields~\cite{Wald1984}. In the vicinity of a compact object, strong gravitational fields can strongly deflect radiation, generating multiple distinct propagating paths that connect a source to an observer~\cite{Perlick2004,VirbhadraEllis2000,Bozza2002,Bozza2010}. For transient or pulsed emission, path-length differences among these trajectories can manifest as characteristic time delays and phase shifts, producing complex echo signals~\cite{ZenginogluGalley2012}. The study of such strong-field multipath propagation has been advanced by breakthroughs in high-resolution interferometry, gravitational-wave detection, and time-domain astronomy~\cite{GibbonsWerner2008,PerlickTsupko2022,LiaoBiesiadaZhu2022}. Observationally, compact emission regions, often modeled as hot spots orbiting near the innermost stable circular orbit of Sagittarius~A*, provide a natural astrophysical laboratory for temporal echo phenomena~\cite{GRAVITY2018,BroderickLoeb2005}. Time-resolved flare imaging further indicates that higher-order image windings can leave secondary peaks in temporal and angular autocorrelations, offering a potential probe of the background geometry~\cite{HadarEtAl2021,ZhangHouGuo2025}.

\par
Despite these astronomical advancements, analyzing repeated winding structures and their coherent time-dependent wave signals in a four-dimensional, dynamically evolving spacetime remains mathematically complex. Motivated by the desire to isolate the geometric mechanisms driving coherent multipath interference, this work addresses a fundamental scientific question: How does a continuous deformation of the underlying intrinsic spatial geometry reconfigure a discrete path-length spectrum and its emergent macroscopic wave response?

\par
To answer this, rather than employing the standard Fermat metric, which describes four-dimensional null geodesics accumulating at the photon sphere \(r_{\rm ph}\), we deliberately investigate the isometric embedding of the constant-time equatorial slice of the RN spacetime~\cite{Lee2012,Poisson2004}. This spatial restriction, generalizing the well-known Schwarzschild Flamm paraboloid to the charged regime, isolates the intrinsic spatial geometry where high-winding spatial geodesics accumulate at the event-horizon throat \(r_+\)~\cite{Flamm2015,EufrasioMecholskyResca2018}. (Appendix~\ref{app:rn-null-geodesics} provides a detailed comparison delineating this distinction from standard four-dimensional RN lensing~\cite{EiroaRomeroTorres2002,HeydariFardEtAl2022}). By coupling this spatial metric to a finite-path surface Huygens--Fresnel construction, we establish a theoretical bridge connecting the discrete geodesic length spectrum to a coherent wave-optics response~\cite{XuWang2021,Ju2026}.

\par
The primary motivation for this specific geometric choice is rooted in the growing field of analogue gravity: in tabletop physical models, such as curved dielectric substrates or metamaterial waveguides, surface waves can be modeled as propagating along the spatial geodesics of the underlying two-dimensional manifold rather than along four-dimensional null trajectories. Unlike the uncharged Schwarzschild limit --- where the analogue geometry is rigidly bound to a single mass scale --- the RN model introduces the black-hole charge parameter as an independent, continuous geometric degree of freedom. At a fixed mass scale, tuning the charge smoothly modifies the horizon radius and reshapes the deep-throat curvature. Consequently, this charge parameter provides a unique mechanism to continuously manipulate the temporal echo intervals and spatial interference nodes without altering the background mass. In the context of transformation optics and laboratory analogues, this theoretical charge tuning can be mapped to concrete experimental design parameters. For instance, replicating this curved geometry on a planar waveguide involves engineering an effective spatially varying refractive-index profile, such as \( n(r) \propto 1/\sqrt{f(r)} \) for radial propagation, or equivalently, manipulating the physical thickness and local curvature of a dielectric substrate~\cite{LeonhardtPhilbin2006,ChenChanSheng2010,BarceloLiberatiVisser2011,Plebanski1960,PendrySchurigSmith2006,Leonhardt2006,GenovZhangZhang2009}.

\par
Recent theoretical and experimental implementations have shown that optical fields, flexural waves, and surface plasmon polaritons can be tightly confined to two-dimensional curved surfaces, where intrinsic curvature acts as a geometric potential~\cite{BatzPeschel2008,SchultheissEtAl2010,DellaValleLonghi2010,BekensteinEtAl2017}. While physical devices must eventually account for material dispersion, polarization, and mode leakage, our geometric framework isolates the geometric contribution to the interference mechanism. It shows how a single geometric parameter controls path deflection, temporal echo intervals, and phase-coherence frequency combs, thereby providing a theoretical blueprint for the design of tunable curved nanophotonic microcavities and analogue-gravity platforms~\cite{ShengEtAl2013}.

\par
The remainder of this paper is structured as follows. Section~\ref{sec:2} derives the exterior RN spatial-slice geometry and its exact isometric embedding in three-dimensional Euclidean space. Section~\ref{sec:multi-loop-winding} formulates the multi loop spatial geodesic boundary value problem and maps the resultant length spectrum to a coherent Huygens-Fresnel wave field. Section~\ref{sec:charge-control} demonstrates how varying the charge shifts the phase alignment spectrum, establishing a parameter inference relation. Section~\ref{sec:5} examines the resulting temporal echoes and spatial interference patterns at a charge selected frequency. Section~\ref{sec.6} summarizes our primary conclusions, while the appendices provide the four-dimensional RN null-geodesic comparison and the supporting analytic and numerical derivations.

\section{Reissner--Nordstr\"om spatial-slice geometry}
\label{sec:2}
\par
In coordinates \(x^\mu=(ct,r,\theta,\phi)\) and with signature \((-,+,+,+)\), the four-dimensional RN metric is~\cite{Poisson2004}
\begin{equation}
ds^2 = -f(r)\,d(ct)^2 + \frac{dr^2}{f(r)} + r^2\left(d\theta^2+\sin^2\theta\,d\phi^2\right),
\label{eq:rn-metric-4d}
\end{equation}
where the metric function \(f(r)\) and the characteristic length scales are given by
\begin{equation}
f(r) = 1 - \frac{r_g}{r} + \frac{r_q^2}{r^2}, \qquad
r_g = \frac{2GM_{\rm BH}}{c^2}, \qquad
r_q^2 = \frac{GQ^2}{4\pi\epsilon_0c^4}.
\label{eq:rn-lapse-function}
\end{equation}
Here \(M_{\rm BH}\) and \(Q\) denote the black hole mass and electric charge, respectively. We assume \(r_q\geq0\), as the geometry depends solely on the magnitude, rather than the sign, of \(Q\). On the constant time equatorial slice \(t=\mathrm{const}\) and \(\theta=\pi/2\), the induced metric in coordinates \(x^a=(r,\phi)\) is~\cite{Lee2012,Poisson2004}
\begin{equation}
h_{ab} =
\begin{pmatrix}
f(r)^{-1} & 0\\
0 & r^2
\end{pmatrix},
\label{eq:rn-metric-2d}
\end{equation}
where the corresponding line element is \(dl^2 = h_{ab}\,dx^a dx^b = f(r)^{-1}dr^2+r^2d\phi^2\). In the full four-dimensional spacetime, \(ds^2\) measures intervals between events. However, once \(t\) and \(\theta\) are fixed, the induced line element \(dl^2\) exclusively measures proper spatial lengths within the slice. In our framework, these lengths serve as analogue optical path lengths, whose differences determine the relative delays and phase shifts among various multipath contributions. The standard null-geodesic problem, which instead relies on the full spacetime line element in Eq.~\eqref{eq:rn-metric-4d}, is summarized separately in Appendix~\ref{app:rn-null-geodesics}.

\par
The constant time embedding must therefore be distinguished from the Fermat metric used to describe four-dimensional null propagation. Here, \(dl\) is an intrinsic slice length and an analogue optical path length. A physical laboratory platform can directly realize this intrinsic spatial geometry. For instance, mapping this metric onto a flat planar waveguide entails engineering an effective refractive-index gradient \(n(r) = 1/\sqrt{f(r)}\). Conversely, lapse, dispersion, and signal loss belong to a separate platform-specific propagation model.

\par
In the nonextremal charged case \(0<r_q<r_g/2\), the function \(f(r)\) has two distinct positive roots~\cite{Poisson2004}:
\begin{equation}
r_\pm = \frac{r_g \pm \sqrt{r_g^2-4r_q^2}}{2} .
\label{eq:rn-horizons}
\end{equation}
The roots \(r_+\) and \(r_-\) correspond to the event and Cauchy horizons, respectively~\cite{Poisson2004}. The geometry recovers the uncharged Schwarzschild case at \(r_q=0\) and becomes extremal at \(r_q=r_g/2\). We restrict our analysis strictly to the exterior region (\(r>r_+\)), where \(f(r)>0\) ensures that \(h_{ab}\) defines a positive-definite spatial metric.

\par
\begin{figure}[tbp]
\centering
\includegraphics[width=5cm]{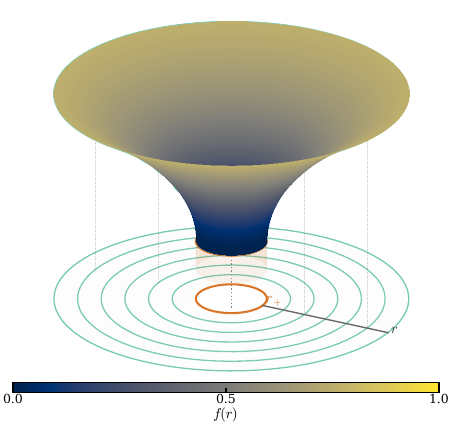}
\caption{Visualization of the isometric embedding of the exterior Reissner-Nordstr\"om spatial slice.}
\label{fig:rn-embedding}
\end{figure}

To visualize this intrinsic geometry, we isometrically embed \(h_{ab}\) into a three-dimensional Euclidean space; Figure~\ref{fig:rn-embedding} shows the resulting surface. It is the charged counterpart of the classical Flamm paraboloid~\cite{Flamm2015,EufrasioMecholskyResca2018}. Unlike the fixed geometric profile of the Schwarzschild limit, the charge parameter \(r_q\) acts as an independent tuning knob, allowing one to continuously elongate the spatial throat while preserving the asymptotic mass scale.

\par
In cylindrical coordinates \((Z,R,\Phi)\), we choose a surface of revolution defined by \(Z=z(r)\), \(R=r\), and \(\Phi=\phi\). Its induced line element is \(dl^2=[1+(dz/dr)^2]dr^2+r^2d\phi^2\). Matching this to \(h_{ab}\) yields
\begin{equation}
\begin{pmatrix}
1+(dz/dr)^2 & 0\\
0 & r^2
\end{pmatrix}
=
\begin{pmatrix}
f(r)^{-1} & 0\\
0 & r^2
\end{pmatrix} .
\label{eq:embedding-match}
\end{equation}
Because the angular components inherently agree, equating the radial components provides the governing embedding equation:
\begin{equation}
\left(\frac{dz}{dr}\right)^2 = \frac{1}{f(r)}-1 .
\label{eq:embedding-ode}
\end{equation}
Throughout the exterior region, the right-hand side is nonnegative, ensuring that the Euclidean embedding remains well defined. For analogue physical implementations, the geometric profile \(z(r)\) provides the exact thickness profile required to fabricate a curved rotationally symmetric substrate.

For \(0<r_q<r_g/2\), the right-hand side of Eq.~\eqref{eq:embedding-ode} diverges as \(r\to r_+^+\), meaning the exterior surface exhibits an infinite slope at its throat. Integrating the positive branch outward with the boundary condition \(z(r_+)=0\) gives
\begin{equation}
z(r) = 2\Bigg[\sqrt{\frac{(r-r_+)\big[(r_++r_-)r-r_+r_-\big]}{r-r_-}} + r_+\big\{F(\alpha(r)\,|\,m)-E(\alpha(r)\,|\,m)\big\}\Bigg] ,
\label{eq:embedding-z}
\end{equation}
where
\begin{equation}
\alpha(r) = \arcsin\sqrt{\frac{r-r_+}{r-r_-}},
\qquad
m = \frac{r_-^2}{r_+^2} .
\label{eq:embedding-z-args}
\end{equation}
Here, \(F(\alpha\,|\,m)\) and \(E(\alpha\,|\,m)\) denote the incomplete elliptic integrals of the first and second kind with parameter \(m\). In the exterior region (\(0<\alpha<\pi/2\) and \(0<m<1\)), Eq.~\eqref{eq:embedding-z} is explicitly real and exactly recovers the positive root of Eq.~\eqref{eq:embedding-ode}. The resulting geometric profile originates at the horizon (\(\lim_{r\to r_+^+}z=0\)) and increases monotonically, with its slope vanishing as \(r\to\infty\) to ensure asymptotic flatness. The limiting cases are detailed below.

\par
At the uncharged endpoint \(r_q=0\), the exterior profile rigorously reduces to the standard Flamm paraboloid~\cite{Flamm2015,EufrasioMecholskyResca2018}:
\begin{equation}
z_{\rm S}(r) = 2\sqrt{r_g(r-r_g)},
\qquad r\geq r_g,
\label{eq:embedding-schwarzschild-limit}
\end{equation}
retaining the finite normalization \(z_{\rm S}(r_g)=0\). At exact extremality, where \(r_+=r_-=r_g/2\), Eq.~\eqref{eq:embedding-ode} simplifies to
\begin{equation}
\frac{dz}{dr} = \frac{\sqrt{r_+(2r-r_+)}}{r-r_+}, \qquad r>r_+ .
\label{eq:embedding-extremal-ode}
\end{equation}
By choosing an exterior reference radius \(r_0>r_+\) and imposing \(z_{\rm ext}(r_0;r_0)=0\), we define the auxiliary functions
\begin{equation}
u(x) = \sqrt{\frac{2x}{r_+}-1}, \qquad
G(x) = 2\sqrt{r_+(2x-r_+)} + r_+\ln\!\left[\frac{u(x)-1}{u(x)+1}\right].
\label{eq:embedding-extremal-primitive}
\end{equation}
These provide the positive branch height difference:
\begin{equation}
z_{\rm ext}(r;r_0) = G(r)-G(r_0), \qquad  r>r_0>r_+ .
\label{eq:embedding-extremal-height-difference}
\end{equation}
Every term remains real because \(u(x)>1\) throughout the exterior, and direct differentiation accurately recovers Eq.~\eqref{eq:embedding-extremal-ode}. However, for any fixed \(r>r_+\),
\begin{equation}
  \lim_{r_0\to r_+^+}z_{\rm ext}(r;r_0) = +\infty .
  \label{eq:embedding-extremal-horizon-limit}
\end{equation}
Thus, the exactly extremal exterior exhibits an infinite embedding-height difference between the horizon and any fixed exterior radius, and consequently cannot be assigned the finite normalization \(z(r_+)=0\).

\par
Along a radial curve with \(d\phi=0\), the radial metric coefficient maps an interval in areal radius to proper radial length through \(dl=dr/\sqrt{f(r)}\). For a fixed exterior areal radius \(R\) and assuming \(r_+>r_-\), the proper length from the event horizon to \(R\) is
\begin{equation}
\ell(R) = \sqrt{(R-r_+)(R-r_-)} + \frac{r_g}{2}\operatorname{arcosh}\!\left(\frac{2R-r_g}{r_+-r_-}\right).
\label{eq:proper-throat-length}
\end{equation}
At a fixed \(R>r_g/2\), the horizon gap \(r_+-r_-=r_g\sqrt{1-4(r_q/r_g)^2}\) narrows and eventually closes as the charge approaches extremality from below. Equation~\eqref{eq:proper-throat-length} then leads to the asymptotic behaviors:
\begin{equation}
\lim_{r_q/r_g\to(1/2)^-}\frac{\ell(R)}{r_g} = +\infty, \qquad
\lim_{r_q/r_g\to(1/2)^-} \frac{\ell(R)/r_g}{-\ln\!\left[1-4(r_q/r_g)^2\right]} = \frac{1}{4}.
\label{eq:near-extreme-throat-divergence}
\end{equation}
Although the interval in areal radius remains finite, its proper radial length diverges, forming an infinitely long spatial throat at extremality. This infinite stretching serves as the equatorial spatial signature of the extremal near-horizon geometry, which locally approaches \(AdS_2\times S^2\)~\cite{KunduriLucietti2013}. Extending the present spatial-geodesic analysis to finite-frequency waves within this elongated throat emerges as a natural next step.

\section{Multipath spatial geodesics and coherent wave field}
\label{sec:multi-loop-winding}
\par
On the constant-time slice, the induced metric determines the admissible multi-loop geodesics and their proper lengths between a localized source and observer. In this section, we identify these spatial paths and translate their discrete length spectrum into a coherent wave field via a finite-path surface Huygens--Fresnel construction~\cite{XuWang2021,Ju2026}. Extending previous Flamm-surface models to the RN geometry~\cite{EufrasioMecholskyResca2018,Ju2026}, this formulation directly maps theoretical high-winding trajectories to physical 2D surface waves, providing a concrete framework for analogue-gravity and transformation-optics experiments.

\par
To evaluate the resulting wave-optical interference~\cite{NambuNoda2016}, we parametrize these geodesics by arc length \(s\), which yields the unit-speed condition~\cite{EufrasioMecholskyResca2018,Ju2026}:
\begin{equation}
f(r)^{-1}\left(\frac{dr}{ds}\right)^2 + r^2\left(\frac{d\phi}{ds}\right)^2 = 1 .
\label{eq:unit-speed}
\end{equation}
Because \(\phi\) is cyclic, axisymmetry yields the conserved geodesic angular-momentum parameter
\begin{equation}
J = r^2\frac{d\phi}{ds}.
\label{eq:conserved-angular-momentum}
\end{equation}
Here, \(J\) is the conserved momentum conjugate to \(\phi\). Let \(\alpha_i\) denote the emission angle at the source radius \(r_i\), measured from the local radial direction. Since the azimuthal component of the initial unit tangent in the local orthonormal frame is simply \(\sin\alpha_i\), evaluating Eq.~\eqref{eq:conserved-angular-momentum} at the source gives
\begin{equation}
J = r_i\sin\alpha_i.
\label{eq:source-angle-angular-momentum}
\end{equation}
This relation explicitly links the local emission geometry to the conserved geodesic label \(J\). Substituting Eq.~\eqref{eq:conserved-angular-momentum} into Eq.~\eqref{eq:unit-speed} to eliminate \(d\phi/ds\) yields the radial equation
\begin{equation}
\left(\frac{dr}{ds}\right)^2 = \frac{(r-r_+)(r-r_-)(r-J)(r+J)}{r^4}.
\label{eq:radial-eq}
\end{equation}
\par
When a geodesic reaches an exterior radial turning point, Eq.~\eqref{eq:radial-eq} dictates \(r_{\rm turn}=J>r_+\). Setting \(\phi_{\rm turn}(J;J)=0\) strategically places the angular origin at that specific turning point. The angular sweep for the turning branch is consequently evaluated as
\begin{subequations}
\label{eq:phi-closed}
\begin{equation}
\phi_{\rm turn}(r;J)= -2J(J-r)(J+r) \times F\!\left(\varphi_{\rm t}(r;J)\,\middle|\,m_{\rm t}\right) \times \big[(J^2-r^2)^2(J+r_-)(J-r_+)\big]^{-1/2},
\label{eq:phi-turning-branch}
\end{equation}
\begin{equation}
\varphi_{\rm t}(r;J)= \operatorname{arccsc}\!\sqrt{\frac{2J(r_--r)}{(J-r)(J+r_-)}}\,, \qquad
m_{\rm t} = \frac{2J(r_--r_+)}{(J+r_-)(J-r_+)} .
\label{eq:phi-turning-amplitude}
\end{equation}
\end{subequations}
Here, \(F(\varphi\,|\,m)\) denotes the incomplete elliptic integral of the first kind, with \(m\) serving as the elliptic parameter following Mathematica's convention.

\par
For the subcritical nonturning branch where \(r>r_0>r_+>r_->0\) and \(0<J<r_+\), the formal root \(r=J\) resides outside the physically accessible exterior domain. Imposing the reference condition \(\phi_{\rm mono}(r_0;r_0,J)=0\) yields the explicitly real solution
\begin{subequations}
\label{eq:phi-mono-subcritical}
\begin{equation}
\phi_{\rm mono}(r;r_0,J) = \frac{2J}{\sqrt{(J+r_-)(r_+-J)}} \Big[ F\!\left(\psi_J(r)\,\middle|\,m_J\right) - F\!\left(\psi_J(r_0)\,\middle|\,m_J\right) \Big],
\label{eq:phi-monotonic-branch}
\end{equation}
\begin{equation}
\psi_J(x) = \arcsin\!\sqrt{\frac{(J+r_-)(x-r_+)}{(x-r_-)(J+r_+)}}\,, \qquad
m_J = \frac{(J-r_-)(J+r_+)}{(J+r_-)(J-r_+)} .
\label{eq:phi-monotonic-amplitude}
\end{equation}
\end{subequations}
The specific endpoint \(J=0\) naturally corresponds to a zero angular sweep. For the nonextremal RN geometry, the opposite endpoint is \(J=r_+\). Employing the same reference condition, its exterior solution takes the form
\begin{subequations}
\label{eq:phi-mono-endpoints}
\begin{equation}
\phi_{\rm mono}(r;r_0,r_+) = \sqrt{\frac{2r_+}{r_+-r_-}} \Big[ F\!\left(\varphi_{r_+}(r_0)\,\middle|\,m_{r_+}\right) - F\!\left(\varphi_{r_+}(r)\,\middle|\,m_{r_+}\right) \Big],
\label{eq:phi-mono-endpoint-branch}
\end{equation}
\begin{equation}
\varphi_{r_+}(x) = \arcsin\!\sqrt{\frac{2r_+(x-r_-)}{(x+r_+)(r_+-r_-)}}\,, \qquad
m_{r_+}=1 .
\label{eq:phi-mono-endpoint-amplitude}
\end{equation}
\end{subequations}
For \(r,r_0>r_+\), the sine arguments in \(\varphi_{r_+}\) exceed unity. Using \(F(\arcsin z\mid 1)=\operatorname{arctanh}z\) on the principal branches, the individual terms can be complex but their imaginary parts cancel in the difference, leaving a strictly real physical angle. These expressions fully describe all exterior branches --- both turning and non-turning --- connecting the source and observer.

\par
For a source and an observer fixed at \(r_i,r_o>r_+\), let \(\Delta\phi_0=|\phi_o-\phi_i|\) be their azimuthal separation. We define \(\Theta \geq 0\) as the total accumulated angle along a trajectory characterized by \(J \geq 0\). Incorporating the \(2\pi\) winding increments from the Flamm-surface finite-path construction~\cite{Ju2026}, the required angular sweeps for the anticlockwise (ACW) and clockwise (CW) families are
\begin{equation}
\Theta_n^{\rm ACW} = \Delta\phi_0 + 2\pi n, \qquad
\Theta_n^{\rm CW} = 2\pi - \Delta\phi_0 + 2\pi n,
\label{eq:multiloop-angles}
\end{equation}
where \(n=0,1,2,\ldots\). For a designated family \(\eta\in\{\mathrm{ACW},\mathrm{CW}\}\), the distinct path characterized by winding order \(n\) is obtained by solving the boundary-value problem:
\begin{equation}
\begin{array}{c}
\displaystyle \frac{dr}{ds} = \sigma_r \sqrt{f(r)\left(1-\frac{J^2}{r^2}\right)}, \qquad \frac{d\Theta}{ds} = \frac{J}{r^2}, \\ \noalign{\vspace{6pt}}
r(0) = r_i, \quad \Theta(0) = 0, \qquad r(S_n) = r_o, \quad \Theta(S_n) = \Theta_n^\eta .
\end{array}
\label{eq:multiloop-bvp}
\end{equation}
Here, the arc length \(s\) originates at the source, and \(\sigma_r=\operatorname{sgn}(dr/ds)\). The terminal conditions determine both the conserved parameter \(J\) and the total spatial path length \(S_n\). A monotonic direct path maintains a constant \(\sigma_r\) and requires \(0\leq J\leq r_{\min}\), where \(r_{\min}=\min\{r_i,r_o\}\). Conversely, a returning path (\(r_+<J\leq r_{\min}\)) contains an interior turning point \(s_t\) where \(r(s_t)=J\) and \(\left.dr/ds\right|_{s=s_t}=0\), reversing the sign of \(\sigma_r\). These two branches meet at \(J=r_{\min}\), where the smaller-radius endpoint itself becomes the turning point. The closed-form angular expressions provide an analytical condition to determine \(J\), from which the total path length \(S_n\) follows directly by evaluating the terminal arc length.

\par
Unlike the purely mass-dependent scaling in Schwarzschild spacetime, the charge parameter shifts the boundary between the direct and returning branches via the modified horizon radii in Eq.~\eqref{eq:radial-eq}. To illustrate this shift, we fix the physical endpoints at \(r_i=4r_g\) and \(r_o=2.4r_g\). At the branch boundary, the smaller endpoint acts as the turning point (\(J=r_o\)), yielding the critical joining angle
\begin{equation}
\Theta_{\rm join} = \int_{r_o}^{r_i} \frac{r_o\,dr}{\sqrt{(r-r_+)(r-r_-)(r^2-r_o^2)}}.
\label{eq:charge-joining-angle}
\end{equation}
This exact expression analytically defines the branch boundary, eliminating the need for numerical root-finding.

\par
\begin{figure}[tbp]
  \centering
  \includegraphics[width=10cm]{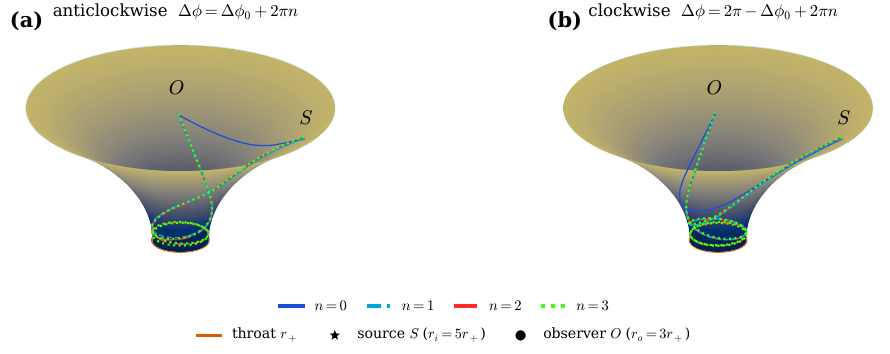}
  \caption{Illustration of multi-loop spatial geodesics on the nonextremal RN analogue surface with \(r_q/r_g=0.4\), \(r_+=0.8r_g\), and \(r_-=0.2r_g\). The source is located at \(r_i=5r_+\), \(\phi_i=0\), and the observer at \(r_o=3r_+\), \(\phi_o=2\pi/3\). Panels (a) and (b) show the anticlockwise (ACW) and clockwise (CW) families, respectively, for \(n=0,1,2,3\). Color and line style consistently identify the winding order. Increasing \(n\) progressively drives the turning point toward the throat and adds one near-throat winding.}
  \label{fig:rn-geodesic-winding}
\end{figure}

Figure~\ref{fig:rn-geodesic-winding} illustrates the two winding families for \(r_i=5r_+\) and \(r_o=3r_+\). For a reference charge \(r_q/r_g=0.4\), Eq.~\eqref{eq:charge-joining-angle} yields \(\Theta_{\rm join}=0.3635\pi\). Since the lowest ACW and CW target sweeps are \(2\pi/3\) and \(4\pi/3\) respectively, all displayed trajectories are returning geodesics. For the primary (\(n=0\)) paths, the conserved parameters are \(J=2.0165r_g\) (ACW) and \(J=1.0736r_g\) (CW). By the third winding order (\(n=3\)), the offset from the spatial throat drops sharply to \(J-r_+=6.89\times10^{-6}r_g\) and \(1.91\times10^{-6}r_g\), respectively. This accumulation of tight inner loops reflects the dynamic turning point \(r=J\) asymptotically approaching the physical horizon \(r=r_+\).

\par
For either winding family, let \(S_n\) denote the total proper length given by Eq.~\eqref{eq:multiloop-bvp}. As the winding order increases, returning geodesics asymptotically approach the throat (\(J_n\to r_+\)), and each additional winding adds exactly one throat circumference:
\begin{equation}
S_{n+1}-S_n \xrightarrow{n\to\infty} 2\pi r_+.
\label{eq:loop-increment}
\end{equation}
Appendix~\ref{app:high-winding-limit} details the derivation of this limit, which is confirmed by the numerical results in Appendix~\ref{app:path-length-convergence} for both orientations.

\par
In the high-winding limit, this throat circumference defines the characteristic temporal echo interval and corresponding resonant frequency for the spatial-slice analogue:
\begin{equation}
\tau_{\rm echo}= \frac{2\pi r_+}{c} = \frac{\pi r_g}{c} \left[1+\sqrt{1-4(r_q/r_g)^2}\right], \qquad
f_{\rm echo}= \frac{1}{\tau_{\rm echo}}.
\label{eq:echo-scale}
\end{equation}
In the Schwarzschild limit (\(r_q=0\)), these reduce to the mass-dependent Flamm-surface echo scales established in Ref.~\cite{Ju2026}. In the RN geometry, the charge provides an additional degree of freedom. At the extremal limit (\(r_q/r_g=1/2\)), the echo time halves, doubling the characteristic echo frequency.

\par
More broadly, classical light rings govern the critical null structure in stationary axisymmetric spacetimes~\cite{CardosoEtAl2009,CunhaHerdeiro2018,PerlickTsupko2022}. In black-hole imaging, near-critical null rays produce a hierarchy of higher-order images accumulating toward the critical curve~\cite{GrallaHolzWald2019,JohnsonEtAl2020,GrallaLupsascaMarrone2020}. For instance, recent ray-tracing analyses of Kerr-Bertotti-Robinson black holes distinguish direct, lensing-ring, and higher-order photon-ring subimages within a synchrotron-emissivity model~\cite{ZhangLiYanYue2026}.

\par
Standard four-dimensional RN null geodesics similarly form clockwise and anticlockwise families with arbitrary windings. However, a key distinction arises between standard RN lensing and our embedded spatial analogue: while standard null rays accumulate at the photon sphere \(r_{\rm ph}\), the spatial geodesics confined to the constant-time slice accumulate at the horizon throat \(r_+\). Consequently, the limiting circumferences differ. The spatial-slice echo scale in Eq.~\eqref{eq:echo-scale} is governed by \(2\pi r_+\), whereas the standard photon-sphere radius and critical impact parameter follow eqs.~\eqref{eq:rn-photon-sphere} and \eqref{eq:rn-critical-impact} (reviewed in Appendix~\ref{app:rn-null-geodesics} using the present notation~\cite{EiroaRomeroTorres2002}).

\par
To relate this localized excitation geometry to an astrophysical source, we model the Gaussian point source as a short-lived flare from a compact hot spot, evaluated at the initial \(t=0\) wavefront. By fixing the source position and pulse envelope at this instant, our approach isolates the coherent propagation from a single wavefront. A continuously orbiting hot spot, which requires a time-dependent source and a retarded superposition of emission events~\cite{BroderickLoeb2005,BroderickLoeb2006,Broderick2006,GRAVITY2018,HadarEtAl2021}, is left for future dynamic extensions of this spatial-slice framework.

\par
To map this discrete spectrum of geometric paths to a macroscopic wave response, we assign a pulsed complex field to each trajectory. Following the Gaussian pulse construction used for Flamm-surface echoes~\cite{Ju2026}, the source incorporates a carrier frequency \(f_0\) and a Gaussian envelope. Setting the temporal pulse center to \(t=0\) yields the initial source field:
\begin{equation}
E_{\rm in}(t) = \exp\!\left[-\frac{t^2}{2\tau^2}\right]\exp(-i\omega_0 t),
\qquad \omega_0 = 2\pi f_0 ,
\label{eq:input-pulse}
\end{equation}
where \(\tau\) represents the standard deviation of the field envelope and the \(1/e\) half-width of the actual intensity envelope. The corresponding physical pulse length is denoted \(\ell_{\rm pulse}=c\tau\). Under the designated Fourier convention \(\tilde{E}(\omega)=\int_{-\infty}^{+\infty}E(t)e^{i\omega t}\,dt\), the source spectrum translates to~\cite{Ju2026}
\begin{equation}
\tilde{E}_{\rm in}(\omega) = \sqrt{2\pi}\,\tau\,\exp\!\left[-\frac{\tau^2}{2}(\omega-\omega_0)^2\right],
\label{eq:input-spectrum}
\end{equation}
featuring a clear center at \(\omega_0\) and a standard spectral deviation \(\Delta\omega=1/\tau\).

\par
\begin{figure}[tbp]
  \centering
  \includegraphics[width=9cm]{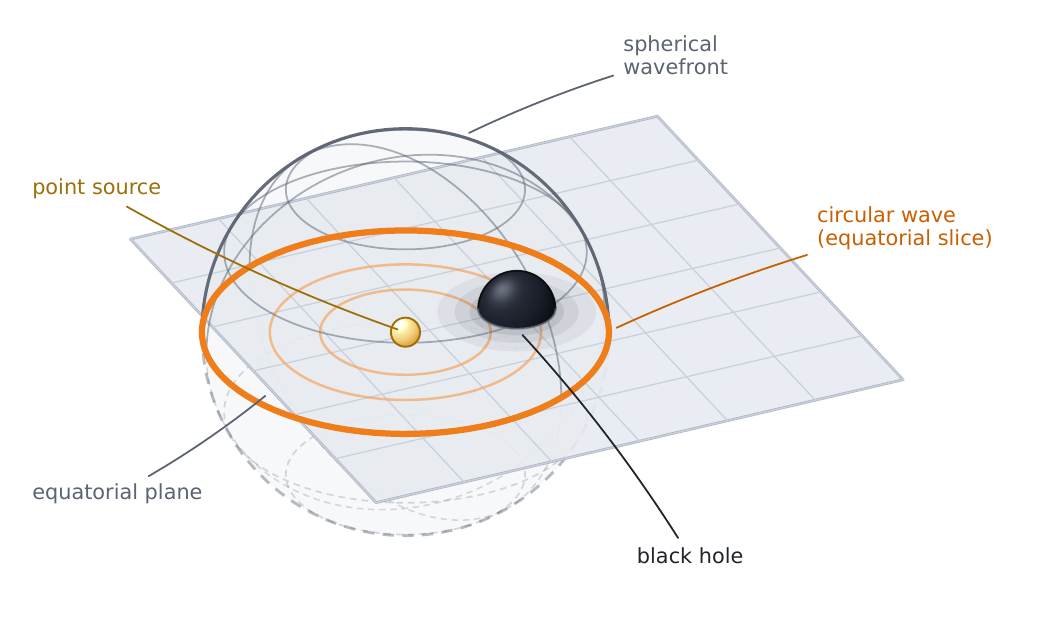}
  \caption{Flat-space point-source reference. A lossless isotropic source inherently emits a spherical wavefront, the equatorial section of which forms a circle. Energy-flux conservation requires \(\displaystyle P=\oint_{\Sigma_L}\langle\mathbf S\rangle\cdot d\mathbf A=4\pi L^2\langle S_r(L)\rangle=\mathrm{const.}\), implying \(\langle S_r(L)\rangle\propto L^{-2}\) and \(|E|\propto L^{-1}\)~\cite{BornWolf2019}. We designate \(1/L\) as the definitive reference path weight, defining \(L\) strictly as the proper geodesic length spanning the RN analogue surface.}
  \label{fig:point-source}
\end{figure}

At \(t=0\), the complete continuous wavefront actively drives the propagation input, with its constituent points acting as distributed secondary Huygens sources. The aggregate complex contributions ultimately reaching the observation location constructively form the net field~\cite{BornWolf2019}. Figure~\ref{fig:point-source} graphically delineates the underlying reference spreading law universally applied to quantify these individual path contributions.

\par
In conventional three-dimensional flat space, an outgoing spherical wave maintains a field amplitude strictly proportional to \(e^{ikL}/L\), where \(k=\omega/c\)~\cite{BornWolf2019}. We preserve the fundamental reference scaling factor \(1/L\) and deliberately assign \(L\) as the exact spatial geodesic length traversing the curved surface. Under the consistent Fourier convention utilized throughout this work, the frequency-domain contribution generated by a single path is expressed as~\cite{XuWang2021,Ju2026}
\begin{equation}
\tilde{E}_{o}(\omega,L) = \frac{\tau}{L}\sqrt{\frac{i\omega}{c}}\,\exp\!\left[-\frac{\tau^2}{2}(\omega-\omega_0)^2\right]e^{ikL}.
\label{eq:single-freq}
\end{equation}
For a strictly narrowband physical pulse satisfying \(\omega_0\tau\gg1\), the energy spectrum remains tightly concentrated near \(\omega_0\). Replacing \(\sqrt{\omega}\) with \(\sqrt{\omega_0}\) within the inverse Fourier transform subsequently provides~\cite{Ju2026}
\begin{equation}
E_o(t,L) \simeq \frac{1}{L}\sqrt{\frac{\omega_0}{2\pi c}}\,\exp\!\left[-\frac{(t-L/c)^2}{2\tau^2}\right] \times \exp\!\left[-i\omega_0(t-L/c)+i\pi/4\right].
\label{eq:single-time}
\end{equation}
For a path of length \(L\), the wave envelope arrives at \(t=L/c\), while the carrier accumulates a phase \(k_0L\), where \(k_0=\omega_0/c\). Consequently, the geodesic length differences determine the relative temporal delays and phase offsets, while a \(1/L\) factor governs the base path amplitudes. Additional effects such as curvature focusing, caustic phase jumps, substrate loss, and finite-aperture coupling can be incorporated into future platform-specific modeling (e.g., customized metamaterial waveguides or shallow water tanks) via a modified propagation amplitude.

\par
The individual path contributions are then superposed to yield the total coherent response. Let \(\mathcal P_{N_{\max}}\) denote the set of geometric paths, comprising the primary direct path and both winding families up to order \(N_{\max}\). Following the finite-path construction of Ref.~\cite{Ju2026} and the present Fourier convention, the total field is given by
{\small
\begin{equation}
\tilde E_{o,T}^{(N_{\max})}(\omega) = \sum_{j\in\mathcal P_{N_{\max}}}\tilde E_o(\omega,L_j), \qquad
E_{o,T}^{(N_{\max})}(t) = \sum_{j\in\mathcal P_{N_{\max}}}E_o(t,L_j), \qquad
I_{o,T}^{(N_{\max})}(t) = \big|E_{o,T}^{(N_{\max})}(t)\big|^2 .
\label{eq:coherent-response}
\end{equation}
}
Equation~\eqref{eq:coherent-response} superposes the complex path amplitudes before calculating the squared magnitude. The resulting cross terms capture wave interference, ensuring the response is a coherent field rather than an incoherent sum of ray intensities. A path of length \(L_j\) introduces a time delay \(L_j/c\) and a phase shift \(2\pi fL_j/c\). Consequently, the geodesic-length difference \(\Delta L_{jk}=L_j-L_k\) acts as the analogue optical-path difference, determining both the relative echo delay \(\Delta L_{jk}/c\) and the interference phase \(2\pi f\Delta L_{jk}/c\).

\par
To isolate the phase interference from the Gaussian envelope and the base \(1/L\) amplitude scaling, we define the weight \(w_j=1/L_j\) and the normalized coefficient \(p_j=w_j/\sum_{k\in\mathcal P_{N_{\max}}}w_k\). The frequency-resolved phase coherence factor is then:
\begin{subequations}
\label{eq:weighted-phase-coherence}
\begin{eqnarray}
&&\mathcal{C}_{N_{\max}}(\phi_o,f) = \left|\sum_{j\in\mathcal P_{N_{\max}}} p_j e^{i2\pi fL_j/c}\right|^2 = \sum_j p_j^2 + 2\sum_{j<k}p_jp_k \cos\!\left(\frac{2\pi f\Delta L_{jk}}{c}\right), \label{eq:weighted-phase-coherence-sum}\\
&&\overline{\mathcal{C}}_{N_{\max}}(f)= \frac{1}{\pi}\int_0^\pi \mathcal{C}_{N_{\max}}(\phi_o,f)\,d\phi_o . \label{eq:weighted-phase-coherence-average}
\end{eqnarray}
\end{subequations}
Each complex exponential acts as a path phasor. Their normalized sum reaches \(\mathcal C_{N_{\max}}=1\) when all phases align, and decreases as they dephase. The pairwise formulation in Eq.~\eqref{eq:weighted-phase-coherence-sum} depends solely on the geodesic-length differences, while the angular average in Eq.~\eqref{eq:weighted-phase-coherence-average} captures phase alignment across observation positions. This quantity represents the squared weighted mean resultant length, analogous to generalized phase-efficiency measures in coherent optical systems~\cite{MardiaJupp2000,AcebronEtAl2005,OlmiBolli2007}.

\par
The lowest-order paths (\(n=0,\ldots,5\)) are obtained from Eq.~\eqref{eq:multiloop-bvp}. At higher winding orders, the offset \(J-r_+\) becomes exponentially small, causing near-throat root-finding to become stiff. We bypass this using the asymptotic relation \(S_n\simeq S_5+(n-5)2\pi r_+\). Appendix~\ref{app:high-winding-limit} derives this approximation, and Table~\ref{tab:rn_geodesic_lengths} confirms its low relative error for both winding families. The macroscopic response sums use \(N_{\max}=100\), which regularizes the weak logarithmic growth inherent in the \(1/L_j\) sum at resonance. The calculations below compare different values of \(N_{\max}\) to show how the interference peaks approach their high-order limits.

\par
In summary, this construction maps the charge-dependent spatial geometry to a coherent wave field. The following sections examine its spectral, temporal, and spatial signatures.

\section{Charge-tuned phase coherence}
\label{sec:charge-control}
\par
In general relativity, a compact object's macroscopic parameters govern wave propagation via the background metric. In our analogue model, the charge parameter \(r_q\) tunes the metric \(f(r)\), controlling the horizon radius \(r_+\), throat circumference, and radial proper distance. These geometric changes alter the multi-loop spatial geodesics and their length spectrum \(\{L_j\}\), whose differences dictate interference phases and echo delays. Building on the geometric analysis of section~\ref{sec:multi-loop-winding}, we now examine how charge controls the coherent wave response, starting in the frequency domain.

\par
The frequency-domain response maps this geometric modulation. Neighboring high-order paths differ asymptotically by one throat circumference (\(2\pi r_+\)), establishing a frequency comb with spacing \(f_{\rm echo}=c/(2\pi r_+)\). At a fixed mass \(r_g\), increasing \(r_q/r_g\) shrinks \(r_+\) and raises \(f_{\rm echo}\) from \(c/(2\pi r_g)\) to the extremal limit \(c/(\pi r_g)\). This charge tunability breaks the mass-frequency lock inherent to the Schwarzschild geometry. By varying the charge, one sweeps this frequency comb across a fixed carrier \(f_0\). When tuned to the resonance condition \(f_0=Mf_{\rm echo}\), adjacent high-order paths differ by \(M\) phase cycles and interfere constructively.

\par
\begin{figure}[tbp]
  \centering
  \includegraphics[width=12cm]{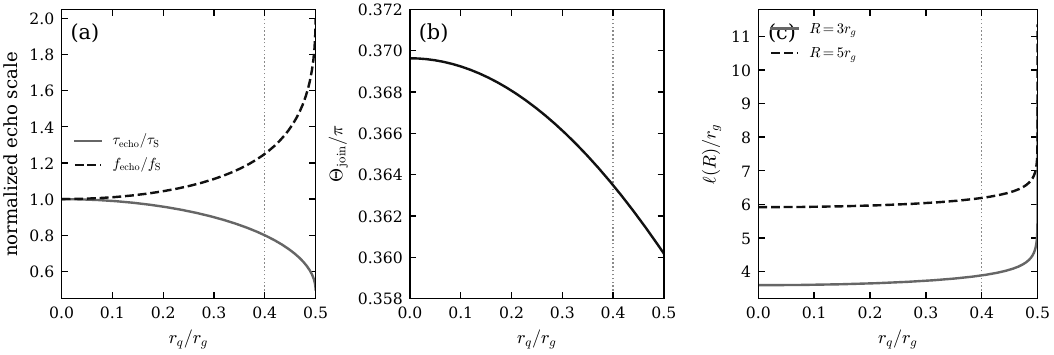}
  \caption{Charge control of the RN spatial-slice analogue at a fixed mass scale \(r_g\). Panels (a)--(c) detail the echo time and frequency normalized by their uncharged Schwarzschild values, the direct--returning joining angle evaluated at \(r_i=4r_g\) and \(r_o=2.4r_g\), and the radial proper distance mapped to \(R=3r_g\) and \(5r_g\), respectively. The vertical dotted line denotes the reference parameter \(r_q/r_g=0.4\).}
  \label{fig:rn-charge-tunability}
\end{figure}

Because the spectral peak spacing is analytically related to the charge, the localized field-point response (Eq.~\eqref{eq:coherent-response}) is best examined first in the frequency domain. Figure~\ref{fig:rn-charge-tunability} establishes the geometric foundation for this response. The vertical dotted line marks \(r_q/r_g=0.4\), serving as the reference geometry for subsequent numerical evaluations. Panel (a) outlines the frequency control range provided by the charge parameter. Panels (b) and (c) demonstrate how this parameter shifts the boundary between path families and alters the radial geometry. Specifically, at \(r_i=4r_g\) and \(r_o=2.4r_g\), the critical joining angle transitions from \(0.3696\pi\) to \(0.3602\pi\) across the charge interval. Notably, the radial proper length diverges as the system approaches extremality. The echo scale, joining angle, and radial proper length all emerge from the RN spatial metric.

\par
For a Gaussian source, the effective bandwidth is \(\Delta f=1/(2\pi\tau)\). Neighboring spatial paths align when their phase difference equals \(2\pi M\). This integer phase-alignment generalizes the echo-frequency resonance of the Flamm surface~\cite{Ju2026}. For either propagation orientation, Eq.~\eqref{eq:loop-increment} yields the alignment condition:
\begin{equation}
f_{M,n} = \frac{Mc}{S_{n+1}-S_n} \xrightarrow[n\to\infty]{} M f_{\rm echo}, \qquad M=1,2,\ldots .
\label{eq:frequency-resonance-condition}
\end{equation}
In the high-winding limit, the resonance frequencies form a spectral comb with uniform spacing \(f_{\rm echo}\), where each line denotes constructive interference across the high-order path family. The multipath response thus develops a sequence of coherent wave maxima, with the finite source bandwidth limiting the number of contributing comb lines. Because low-order paths have nonuniform length increments, their corresponding maxima shift away from the asymptotic sequence. While the spatial geometry fixes the theoretical peak positions via path differences, physical realizations (e.g., a metamaterial waveguide) will introduce path-dependent coupling efficiencies and propagation losses that determine the measured peak amplitudes.

\par
Under the ideal reference kernel \(G_0(\omega,L)\propto e^{ikL}/L\), the spatial geometry dictates the peak positions. However, a laboratory platform (such as a planar waveguide or acoustic resonator) replaces \(G_0\) with a platform-specific Green function \(G_{\rm plat}(\omega;\mathbf{x}_o,\mathbf{x}_s)\). Its phase accumulation is governed by a propagation constant \(\beta(\omega)\), while its amplitude incorporates polarization, material loss, cross-coupling, and radiation leakage~\cite{Plebanski1960,Yariv1973,Marcuse1976,KristensenQNM2020}. Thus, while \(f_{\rm echo}\) remains a universal geometric spacing, platform-specific group delays and line shapes depend on the physical propagation law (see Appendix~\ref{app:platform-green}). These phase-alignment peaks should not be confused with a device's global eigenmode resonances unless such a correspondence is established by a platform-specific wave equation. Equation~\eqref{eq:weighted-phase-coherence} separates this geometric phase organization from the Gaussian spectral envelope and the reference \(1/L\) amplitudes.

\par
\begin{figure}[tbp]
  \centering
  \includegraphics[width=12cm]{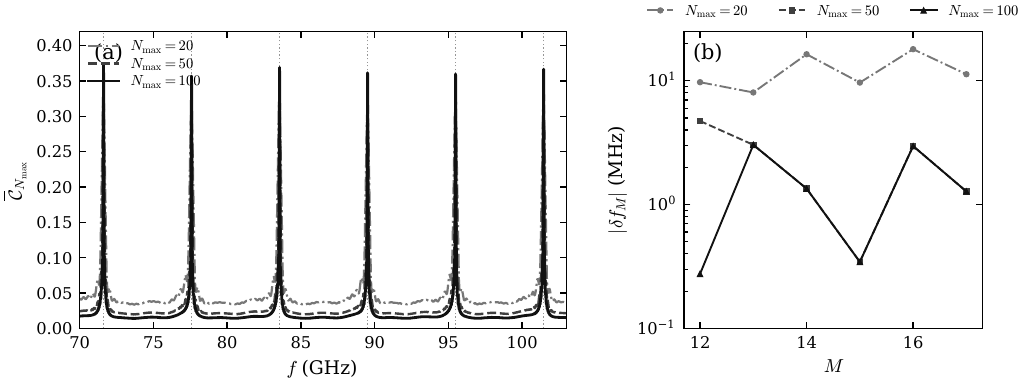}
  \caption{Angle-averaged weighted phase coherence at \(r_q/r_g=0.4\), \(r_i=4r_g\), and \(r_o=6.4r_g\). The dash-dotted, dashed, and solid curves utilize \(N_{\max}=20,50,100\), while the dotted vertical lines mark \(f=Mf_{\rm echo}\) for \(M=12,\ldots,17\). The interference peaks approach the predicted comb spacing as \(N_{\max}\) increases.}
  \label{fig:rn-phase-coherence-spectrum}
\end{figure}

Figure~\ref{fig:rn-phase-coherence-spectrum} maps the frequency-domain response corresponding to the geometric parameters in Figure~\ref{fig:rn-charge-tunability}, using \(r_q/r_g=0.4\), \(r_i=4r_g\), and \(r_o=6.4r_g\). For harmonic orders \(M=12,\ldots,17\), the largest frequency offset from the theoretical \(Mf_{\rm echo}\) decreases from \(17.97\,\mathrm{MHz}\) at \(N_{\max}=20\) to \(3.03\,\mathrm{MHz}\) at \(N_{\max}=100\). This asymptotic offset falls below the \(5\,\mathrm{MHz}\) numerical sampling interval. The stabilized peak centers recover the analytical relation \(r_q\mapsto r_+\mapsto f_{\rm echo}\), confirming that \(f_{\rm echo}\) is the charge-controlled frequency comb spacing.

\par
For a fixed carrier frequency \(f_0\), the charge dependence of \(r_+\) establishes the phase-matching ratio:
\begin{equation}
\frac{f_0}{f_{\rm echo}} = \frac{\pi r_g f_0}{c} \left[1+\sqrt{1-4(r_q/r_g)^2}\right].
\label{eq:fixed-carrier-charge-ratio}
\end{equation}
Constructive interference occurs when \(f_0/f_{\rm echo}=M\). Let \(r_q^{(M)}\) denote the charge parameter required to align the \(M\)th comb line with \(f_0\). Solving this matching condition yields:
\begin{equation}
\frac{r_q^{(M)}}{r_g} = \frac{1}{2}\left[ \frac{M c}{\pi r_g f_0} \left(2-\frac{M c}{\pi r_g f_0}\right) \right]^{1/2},
\label{eq:fixed-carrier-resonant-charge}
\end{equation}
where the available mode numbers are bounded by \(\pi r_g f_0/c\leq M\leq2\pi r_g f_0/c\). This equation provides the charge-tuning rule: for a given carrier, one selects an integer \(M\) in this domain, and Eq.~\eqref{eq:fixed-carrier-resonant-charge} determines the analogue charge required to lock the \(M\)th comb line at \(f_0\). Because increasing the charge raises the comb spacing, the matched integer mode \(M\) must decrease. This forward mapping parallels four-dimensional astrophysical strong lensing, where relative delays and image structures encode the lensing geometry~\cite{LiaoBiesiadaZhu2022}.

\par
Conversely, if neighboring high-order spectral peaks are resolved with a common spacing \(\Delta f_{\rm comb}\simeq f_{\rm echo}\), a known mass scale \(r_g\) yields:
\begin{equation}
\frac{r_q}{r_g} = \frac{1}{2}\left[ \frac{c}{\pi r_g\Delta f_{\rm comb}} \left(2-\frac{c}{\pi r_g\Delta f_{\rm comb}}\right) \right]^{1/2}.
\label{eq:charge-inversion-from-comb}
\end{equation}
Here, the observable spacing is bounded by \(c/(2\pi r_g)\leq\Delta f_{\rm comb}\leq c/(\pi r_g)\). Equation~\eqref{eq:charge-inversion-from-comb} inverts a spectral spacing into an estimate of the effective charge. Resolving several neighboring peaks determines their common spacing and separates it from finite-order peak shifts. Subsequent calibration of wave coupling, loss, and frequency uncertainty connects this geometric estimate to a physical device. Thus, in analogue gravity, a wideband frequency scan can verify a prescribed analogue geometry or infer the effective charge of an unknown sample.

\par
At resonance (\(f=Mf_{\rm echo}\)), the high-order anticlockwise and clockwise wave contributions carry angular phases of opposite signs. Their interference yields the RN-slice counterpart of the angular-node relation from Ref.~\cite{Ju2026}:
\begin{equation}
\widetilde I_M(\phi_o) \propto \cos^2(M\phi_o), \qquad
\phi_{o,m} = \frac{(m+1/2)\pi}{M}, \qquad m=0,\ldots,M-1.
\label{eq:frequency-angular-nodes}
\end{equation}
The resulting high-order wave pattern contains \(M\) angular nodes. Including low-order spatial paths modifies the node depths but preserves this angular organization.

\par
\begin{figure}[tbp]
\centering
\includegraphics[width=12cm]{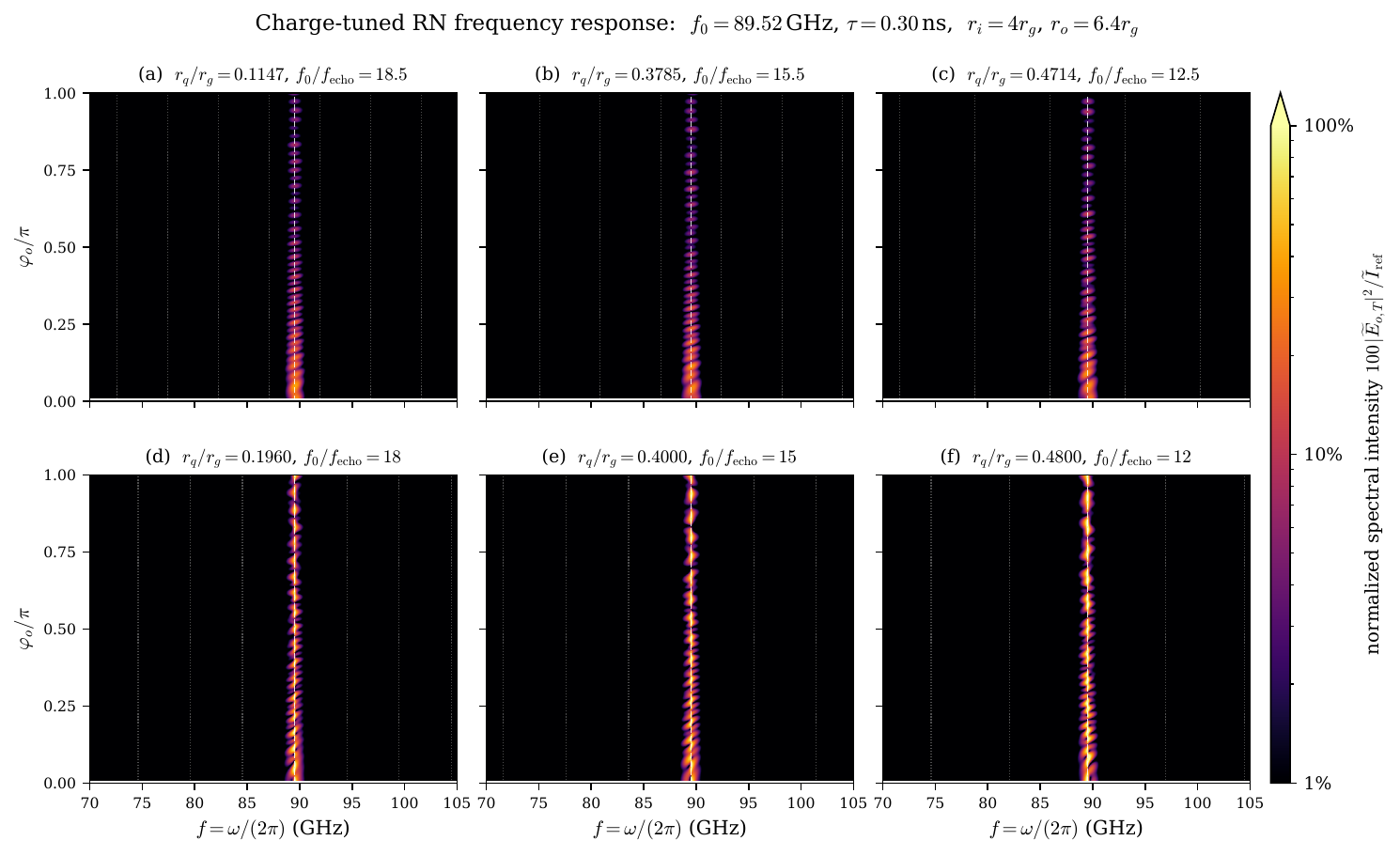}
\caption{Charge matching evaluated at a carrier frequency \(f_0=89.5247\,{\rm GHz}\) using \(N_{\max}=100\), \(\tau=0.30\,{\rm ns}\), \(r_i=4r_g\), and \(r_o=6.4r_g\). The respective data columns correspond to \(M=18,15,12\); the upper and lower rows enforce the tuning relations \(f_0/f_{\rm echo}=M+1/2\) and \(M\), respectively. The vertical lines mark \(f=Mf_{\rm echo}\) and the target carrier \(f_0\). All panels share the globally fixed intensity reference \(\widetilde I_{\rm ref}=2\pi\tau^2I_{\rm ref}\) derived at \(r_q/r_g=0.4\).}
\label{fig:rn-frequency-response}
\end{figure}

To compare disparate charge values on a fixed color scale, we eliminate the \(\tau^2\) factor from the Fourier transform by defining:
\begin{equation}
\widetilde I_{\rm ref}(\tau,f_0) = 2\pi\tau^2 I_{\rm ref}(f_0).
\label{eq:spectral-intensity-reference}
\end{equation}
For Figure~\ref{fig:rn-frequency-response}, \(I_{\rm ref}\) is evaluated once at \(r_q/r_g=0.4\) and held fixed. The normalized quantity \(100|\widetilde E_{o,T}^{(N_{\max})}|^2/\widetilde I_{\rm ref}\) places all panels onto a unified intensity scale, making their charge dependence comparable.

\par
Figure~\ref{fig:rn-frequency-response} evaluates this analytical test at \(f_0=89.5247\,{\rm GHz}\), keeping the carrier, pulse width, and positions fixed across all panels. The upper row applies a half-integer phase detuning (\(f_0/f_{\rm echo}=M+1/2\)), whereas the lower row uses \(r_q/r_g=0.195959, 0.400000, 0.480000\). These parameters satisfy the matching condition in Eq.~\eqref{eq:fixed-carrier-resonant-charge} for \(M=18,15,12\), respectively. The phase-matched cases demonstrate enhanced constructive interference at the target carrier. Thus, varying \(r_q/r_g\) alone translates a selected comb line onto a prescribed operating frequency.

\par
Figures~\ref{fig:rn-charge-tunability} through \ref{fig:rn-frequency-response} establish the mapping from geometric parameters to macroscopic wave response in this model. The charge parameter modifies the spatial metric, which alters the path-length spectrum and determines the frequency comb. This comb can be tuned to a chosen carrier frequency or inverted to estimate the effective charge parameter. In a laboratory analogue, this parameter serves as a design and calibration variable. A four-dimensional extension of this work would replace the spatial lengths \(L_j\) with relativistic null-geodesic travel times and parallel-transport amplitudes, or use a full-wave covariant solution. In such an extension, the resulting spectra, echoes, and multipath images could be examined jointly as multiphysics probes of compact-object parameters.

\section{Spatiotemporal echoes and interference fringes}
\label{sec:5}
\par
To examine this phase aligned state in time and space, we fix the reference geometry at \(r_q/r_g=0.4\). At this charge, the carrier \(f_0=89.52\,{\rm GHz}\) equals \(15f_{\rm echo}\), yielding \(M=15\). The associated spectral peak indicates that the phase difference between neighboring high-order paths approaches \(2\pi M\), leading to constructive interference. As shown in Figure~\ref{fig:rn-phase-coherence-spectrum}, the coherence maxima coincide with the predicted frequency comb. In Figure~\ref{fig:rn-frequency-response}, charge matching produces a stronger carrier-centered ridge than half-integer detuning. These peaks mark constructive multipath interference organized by the RN spatial geometry. They are phase-alignment features of the finite reference path sum, distinct from device eigenmode resonances unless a platform-specific wave equation establishes otherwise.

\par
\begin{figure}[tbp]
  \centering
  \includegraphics[width=12cm]{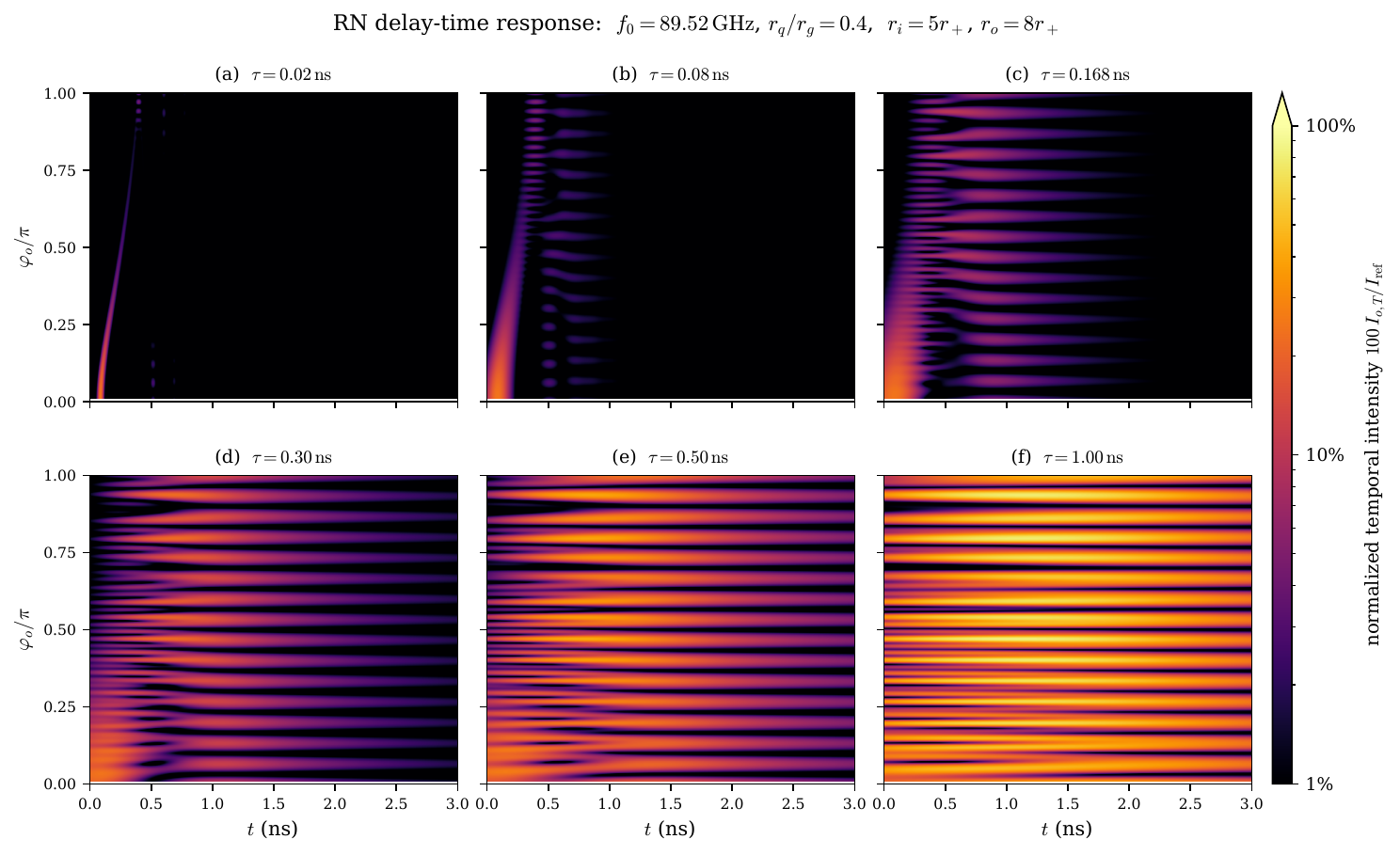}
  \caption{Transient delay-time response for \(M=15\) at \(f_0=15f_{\rm echo}\). Increasing the pulse duration ratio \(\tau/\tau_{\rm echo}\) transforms the isolated primary arrival and discrete echoes in panels (a) and (b) into the connected resonant ridges in panel (c), and eventually into the continuous coherent tails in panels (d)--(f). The alternating angular bands show that the angular phase structure remains visible even after neighboring arrivals overlap. The common intensity normalization allows the \(t=0.30\,{\rm ns}\) temporal section to be compared directly with the \(r=8r_+\) spatial circle in Figure~\ref{fig:rn-energy-snapshot}.}
  \label{fig:rn-delay-time}
\end{figure}
\begin{figure}[tbp]
  \centering
  \includegraphics[width=12cm]{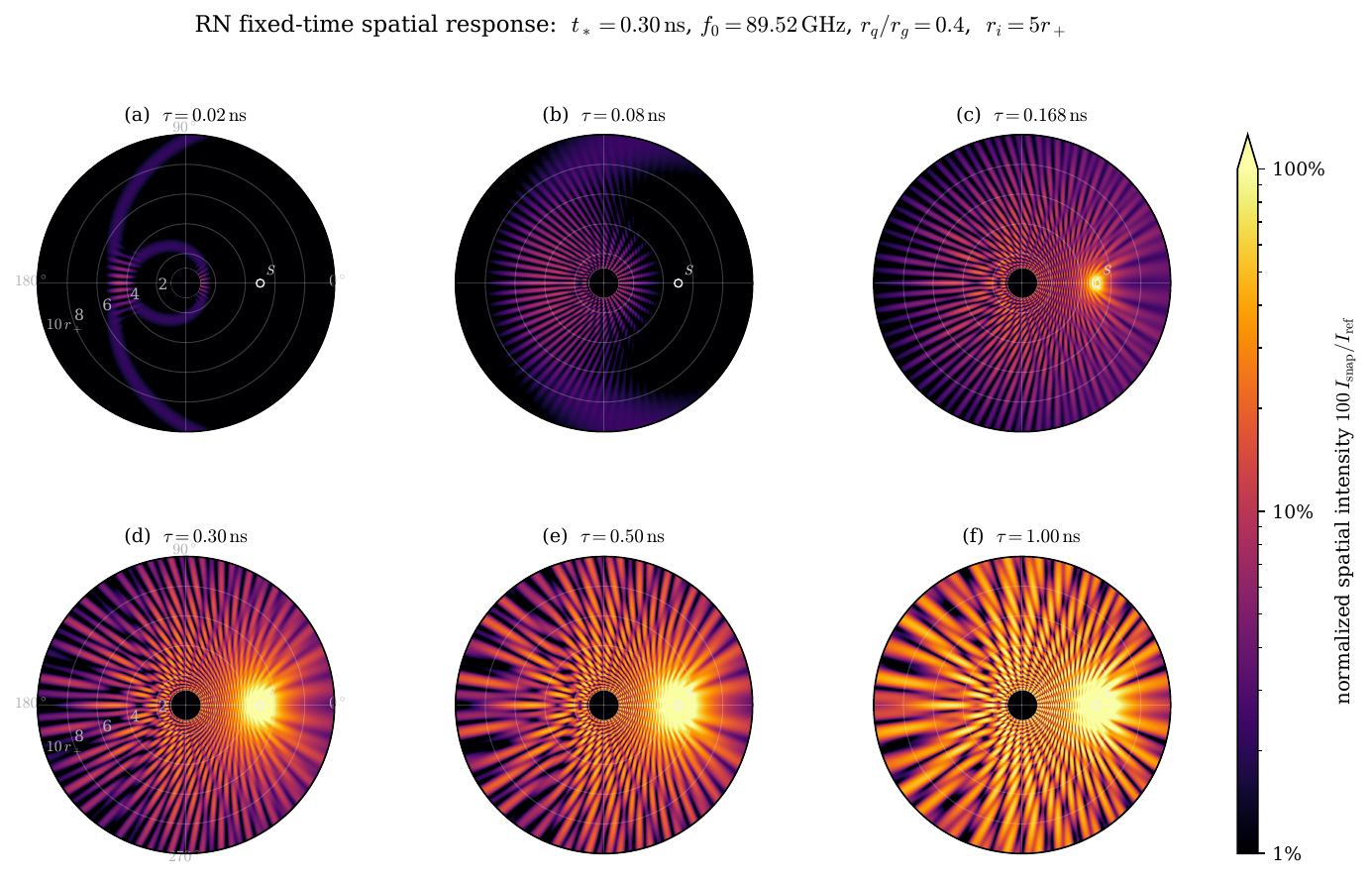}
  \caption{Instantaneous fixed-time spatial wave response for \(M=15\), corresponding to Figure~\ref{fig:rn-delay-time}. A short temporal pulse isolates discrete equal-arrival wavefronts in panel (a). As \(\tau/\tau_{\rm echo}\) increases, overlapping path families form the radial and angular interference network in panels (b)--(f). The high-order spatial fringes crowd into dense concentric rings near the throat and open into broader angular fans at larger radii. The blank central disk corresponds to the inaccessible interior region, and \(S\) marks the point source. The \(r=8r_+\) circular contour indicates the observation section matching the delay-time response.}
  \label{fig:rn-energy-snapshot}
\end{figure}

Using the same \(M=15\) path spectrum and a common reference intensity \(I_{\rm ref}=4I_{\rm dir}^{\rm peak}\) (where \(I_{\rm dir}^{\rm peak}\) is the peak direct-path intensity), Figures~\ref{fig:rn-delay-time} and \ref{fig:rn-energy-snapshot} map the temporal and spatial responses. This shared normalization allows direct color comparison between the two figures. The relevant geometric scales are \(\tau_{\rm echo}=0.168\,{\rm ns}\) and \(f_{\rm echo}=5.97\,{\rm GHz}\).

\par
Temporal crossover is governed by \(\tau/\tau_{\rm echo}\). The pulse length \(c\tau\) sets the longitudinal extent of each arrival, while \(c\tau_{\rm echo}=2\pi r_+\) is the asymptotic separation between neighboring high-order paths. For \(\tau\ll\tau_{\rm echo}\), these length windows do not overlap. Figure~\ref{fig:rn-delay-time}(a) shows a curved first-arrival front --- reflecting the variation of the shortest path with observer angle --- followed by isolated higher-winding features. These later arrivals form a discrete echo ladder as successive winding orders remain resolved. In panel (b), adjacent envelopes begin to overlap, though their maxima remain distinguishable.

\par
As \(\tau\) approaches \(\tau_{\rm echo}\), panel (c) shows neighboring arrivals merging into extended ridges. For the broader pulses in panels (d)--(f), multiple winding orders overlap in time, transforming the discrete ladder into a continuous multipath tail. This continuity preserves the phase information. At \(f_0=15f_{\rm echo}\), the asymptotic phase increment is \(2\pi M\), maintaining constructive interference among high-order paths. The ACW and CW families carry opposite angular phases, producing a regular pattern of alternating bright and dark bands across \(\phi_o\) described by Eq.~\eqref{eq:frequency-angular-nodes}. This \(M=15\) angular organization persists even when individual arrivals overlap, yielding a continuous, phase structured tail rather than a simple pulse broadening.

\par
While the temporal response captures arrival times, a fixed-time spatial snapshot reveals where these paths interfere. For any exterior point \((r,\phi)\), the boundary-value problem provides the path spectrum \(S_j(r,\phi)\), yielding the snapshot:
\begin{equation}
I_{\rm snap}^{(N_{\max})}(r,\phi;t_*)=\left|\sum_{j\in\mathcal P_{N_{\max}}}E_o\!\left(t_*,S_j(r,\phi)\right)\right|^2.
\label{eq:spatial-snapshot}
\end{equation}
At a fixed time \(t_*\), the Gaussian factor in Eq.~\eqref{eq:single-time} limits the dominant paths to \(\lvert S_j-ct_*\rvert\lesssim c\tau\). A short pulse thus selects a narrow shell in path-length space, displayed in Figure~\ref{fig:rn-energy-snapshot}(a) as separated equal-arrival wavefronts. Increasing \(\tau\) widens this length window, allowing direct, ACW, and CW branches to coexist and interfere at each point. The progression across panels (b)--(f) demonstrates this transition from isolated fronts to an extended spatial interference network.

\par
The spatial interference exhibits distinct radial and angular structures. Radially, high-winding solutions satisfy \(J\to r_+^+\). Their turning points and equal-phase contours crowd near the throat, producing dense inner rings. Farther out, path lengths vary more gradually, opening the fringes into broader radial fans. Angularly, the ACW and CW phases generate the \(M\)-node pattern of Eq.~\eqref{eq:frequency-angular-nodes} over \(0\leq\phi\leq\pi\), repeated by reflection symmetry across the full domain. While direct and low-order paths modulate the fringe contrast and curvature, the high-order sector strictly retains \(M=15\). The angular node count thus serves as a spatial signature of the integer defining the spectral comb. The response remains continuous across the direct-returning joining angle, as both branches share the same angular sweep and path length at \(J=r_{\min}\).

\par
Unlike four-dimensional photon and lensing rings governed by null orbits approaching \(r_{\rm ph}\), the near-throat rings in Figure~\ref{fig:rn-energy-snapshot} are governed by spatial geodesics approaching \(r_+\). Appendix~\ref{app:rn-null-geodesics} summarizes the underlying null-geodesic relations~\cite{CunhaHerdeiro2018,GrallaHolzWald2019}. This distinction stems from the different conserved structures of the two geodesic problems. For equatorial null motion in four-dimensional RN spacetime, the time and axial Killing fields provide conserved quantities \(\mathcal E_\gamma\) and \(\mathcal L_\gamma\), whose ratio defines the impact parameter \(b_{\rm null}=\mathcal L_\gamma/\mathcal E_\gamma\) (Eq.~\eqref{eq:null-first-integrals}). In contrast, the constant-time induced metric lacks a temporal direction; its unit-speed spatial geodesics retain only the angular constant \(J\). Thus, four-dimensional high-winding null paths are organized by the critical impact parameter and \(r_{\rm ph}\), whereas these spatial paths are organized by \(J\to r_+^+\).

\par
A related separation between geometry and radiative modeling also appears in four-dimensional imaging. The background metric sets the critical photon-orbit scale, whereas the brightness and visibility of higher-order rings depend on the accretion prescription and angular resolution~\cite{ZengEtAl2022}.

On the observer circle, the temporal and spatial responses are linked point by point:
\begin{equation}
I_{\rm snap}^{(N_{\max})}(r_o,\phi;t_*) =I_{o,T}^{(N_{\max})}(\phi,t_*).
  \label{eq:time-space-section}
\end{equation}
This circle is simply the fixed-time section of the delay-time map. This common section provides an internal consistency check, connecting each angular fringe at \(r_o\) directly to its temporal counterpart.

Figures~\ref{fig:rn-delay-time} and \ref{fig:rn-energy-snapshot} compare the temporal and spatial responses at the same charge-selected spectral peak. The delay-time response captures the transition from discrete echoes to a continuous coherent tail, while the fixed-time response shows localized fronts evolving into extended radial and angular fringes. Frequency, time, and space thus provide three complementary projections of a single charge-dependent path-length spectrum: the peak spacing identifies \(f_{\rm echo}\), the echo overlap reflects \(\tau/\tau_{\rm echo}\), and the angular nodes encode the integer \(M\).

\section{Conclusion}
\label{sec.6}
\par
In this work, we established a rigorous theoretical mapping between the intrinsic spatial geometry of the RN spacetime slice and the coherent macroscopic response of a wave field. By representing the exterior constant-time equatorial slice as an isometrically embedded curved surface and employing a finite-path surface Huygens--Fresnel framework, we demonstrated how the analogue charge parameter explicitly reorganizes the multipath length spectrum. Solving the exact geodesic boundary-value problem identifies one direct branch and two oriented returning families. Coherent superposition of these branches directly translates spatial path lengths into measurable phase delays and interference fringes. Consequently, the charge parameter acts as a fundamental physical dial: it modifies the background metric, dictates the discrete path-length sequence, and ultimately governs the coherent wave response.

\par
In the high-winding limit, the throat circumference fundamentally dictates the multipath sequence. As the turning radius of high-order paths approaches the horizon \(r_+\), the length increment between successive windings strictly converges to \(2\pi r_+\). This geometric limit locks the echo time \(\tau_{\rm echo}\) and the fundamental comb spacing \(f_{\rm echo}\). At a constant mass \(r_g\), tuning the charge parameter from zero to the extremal limit smoothly halves the echo time and doubles the frequency comb spacing. Simultaneously, this charge variation shifts the spatial boundary separating direct and returning waves, while the radial proper distance to the throat diverges near extremality. The exterior spatial slice thus possesses two distinct, tunable geometric scales: a finite azimuthal winding scale tied to the horizon, and a radial distance scale that grows without bound as the system approaches extremality.

\par
When this discrete path spectrum is promoted to a coherent wave field, the underlying geometric scales uniquely manifest across the frequency, time, and space domains. In the frequency domain, the coherent response forms a spectral comb \(f=Mf_{\rm echo}\). By varying the analogue charge, this comb can be actively swept across a fixed carrier frequency to achieve perfect phase resonance; conversely, measuring this comb spacing provides a direct inversion to infer the effective charge of an unknown geometry. In the time domain, the ratio of the pulse width to the echo time (\(\tau/\tau_{\rm echo}\)) governs the crossover from a ladder of isolated discrete echoes to a continuous, phase-coherent tail. In the spatial domain, this identical path superposition yields dense radial rings and structured angular interference fringes. Therefore, the spectral comb, the transient echo ladder, and the steady-state spatial fringes are not independent phenomena, but rather three complementary physical projections of a single charge-controlled path-length spectrum.
\par
\begin{figure}[tbp]
  \centering
  \includegraphics[width=7cm]{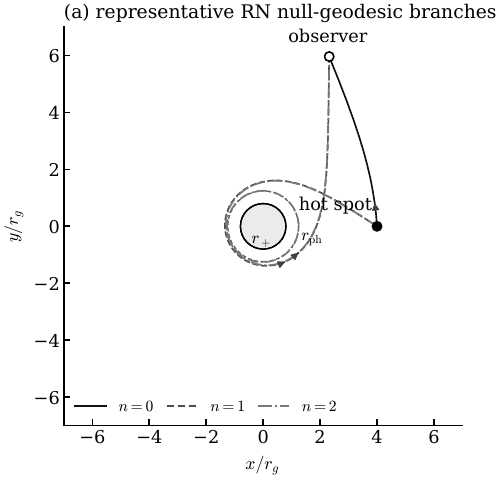}
  \caption{Representative null-geodesic branches in a static four-dimensional RN background with a compact hot-spot source. Higher-winding branches approach the photon sphere \(r_{\rm ph}\), illustrating the necessary geometric input for a future four-dimensional extension of the present path-based response framework.}
  \label{fig:rn-4d-hotspot-geometry}
\end{figure}

The modularity of this theoretical framework makes it highly extensible. Because the spatial metric strictly dictates the geometric path spectrum, platform-specific effects --- such as material dispersion, propagation loss, and mode coupling --- can be independently incorporated through a customized Green function. This ensures the effective analogue charge remains a continuous design variable for engineering interference in curved metamaterial waveguides or surface-wave resonators. Furthermore, this framework conceptually paves the way for four-dimensional astrophysical extensions. In a full spacetime context, the spatial lengths \(L_j\) would naturally be replaced by relativistic null-geodesic travel times \(T_j\), while the scalar spreading weights would be upgraded to transport amplitudes encompassing gravitational redshift, lensing magnification, and polarization rotation. As illustrated in Figure~\ref{fig:rn-4d-hotspot-geometry} for a compact hot-spot source, higher-winding optical branches would systematically accumulate at the photon sphere rather than the spatial throat. Despite this shift in the organizing critical geometry, the core methodology --- utilizing a discrete set of winding paths to synthetically construct coherent spectra, macroscopic echoes, and multipath images --- remains robust, offering a unified multiphysics approach to probing the macroscopic parameters of relativistic compact objects.

\appendix

\section{Four-dimensional RN null geodesics}
\label{app:rn-null-geodesics}
\par
This appendix collects the standard null geodesics of the four-dimensional RN spacetime in Eq.~\eqref{eq:rn-metric-4d}. Together with the spatial-geodesic results in the main text, these equations show how similar winding families are organized by different critical geometries.

\par
Restrict the motion to the equatorial plane, set \(x^0=ct\), and parameterize the trajectory by an affine parameter \(\lambda\). The null condition and the first integrals associated with stationarity and axisymmetry are
\begin{eqnarray}
&&0 = \frac{1}{2}\Bigg[-f(r)\left(\frac{d x^0}{d\lambda}\right)^2+\frac{1}{f(r)}\left(\frac{d r}{d\lambda}\right)^2+r^2\left(\frac{d\phi}{d\lambda}\right)^2\Bigg],
\label{eq:null-geodesic-lagrangian} \\
&&\mathcal E_\gamma = f(r)\frac{d x^0}{d\lambda},
\qquad
\mathcal L_\gamma=r^2\frac{d\phi}{d\lambda} .
\label{eq:null-first-integrals}
\end{eqnarray}
Here \(\mathcal E_\gamma\) and \(\mathcal L_\gamma\) are the conserved photon energy and angular momentum in the chosen affine parametrization. These equations represent the RN limit of the standard static spherical null-geodesic equations collected in Ref.~\cite{HeydariFardEtAl2022}.

\par
The affine-rescaling-invariant impact parameter is
\begin{equation}
b_{\rm null}=\frac{\mathcal L_\gamma}{\mathcal E_\gamma}.
\label{eq:null-impact-parameter}
\end{equation}
The radial equation and the effective potential then take the form
\begin{equation}
\left(\frac{d r}{d\lambda}\right)^2+f(r)\frac{\mathcal L_\gamma^2}{r^2} = \mathcal E_\gamma^2, \qquad
V_{\rm eff}(r) = f(r)\frac{\mathcal L_\gamma^2}{r^2}.
\label{eq:null-radial-effective-potential}
\end{equation}
Eliminating the affine parameter gives the orbit equations
\begin{eqnarray}
&&\left(\frac{dr}{d\phi}\right)^2= \frac{r^4}{b_{\rm null}^2}-r^2f(r),
\label{eq:null-orbit-r}\\
&&\left(\frac{du}{d\phi}\right)^2= \frac{1}{b_{\rm null}^2}-u^2+r_g u^3-r_q^2u^4,~~ u=\frac{1}{r}.
\label{eq:null-orbit-u}
\end{eqnarray}
Equations~\eqref{eq:null-impact-parameter}--\eqref{eq:null-orbit-u} follow from the null-geodesic equations in Ref.~\cite{HeydariFardEtAl2022} after specializing its metric functions to the RN case.

\par
For a ray that comes from infinity and returns to infinity, let \(r_0\) be the radius of closest approach. Setting \(d r/d\lambda=0\) at \(r_0\) gives
\begin{equation}
b_{\rm null}(r_0)=\frac{r_0}{\sqrt{f(r_0)}}.
\label{eq:null-turning-impact-parameter}
\end{equation}
The corresponding deflection angle is
\begin{equation}
\widehat\alpha(r_0)=2\int_{r_0}^{\infty}\frac{dr}{r\sqrt{(r/r_0)^2f(r_0)-f(r)}}-\pi .
\label{eq:null-deflection-angle}
\end{equation}
These RN impact parameter and deflection formulas match those in Ref.~\cite{EiroaRomeroTorres2002}, adapted to the present lapse function and length scales.

\par
An unstable circular null orbit occurs at the exterior maximum of \(f(r)/r^2\). Its physical outer root is
\begin{equation}
r_{\rm ph}= \frac{3r_g+\sqrt{9r_g^2-32r_q^2}}{4}.
\label{eq:rn-photon-sphere}
\end{equation}
The associated critical impact parameter is
\begin{equation}
b_{\rm c}
=\frac{r_{\rm ph}}{\sqrt{f(r_{\rm ph})}}.
\label{eq:rn-critical-impact}
\end{equation}
The photon-sphere radius in Eq.~\eqref{eq:rn-photon-sphere} and the critical impact parameter in Eq.~\eqref{eq:rn-critical-impact} are the standard RN results~\cite{EiroaRomeroTorres2002,HeydariFardEtAl2022}. They reduce to \(r_{\rm ph}=3r_g/2\) and \(b_{\rm c}=3\sqrt{3}r_g/2\) in the Schwarzschild limit. At the extremal endpoint, they become \(r_{\rm ph}=r_g\) and \(b_{\rm c}=2r_g\).

\par
For rays incident from infinity, \(b_{\rm null}<b_{\rm c}\) corresponds to capture, \(b_{\rm null}=b_{\rm c}\) asymptotically approaches the unstable circular orbit, and \(b_{\rm null}>b_{\rm c}\) yields a scattering orbit with an exterior turning point. The deflection angle diverges as \(r_0\to r_{\rm ph}^{+}\), producing null trajectories with arbitrarily many windings and two oriented families of relativistic images~\cite{EiroaRomeroTorres2002,HeydariFardEtAl2022}. Thus, four-dimensional null trajectories accumulate at the photon-sphere radius \(r_{\rm ph}\), which lies outside the event horizon. In contrast, the spatial geodesics in the main text accumulate at the throat radius \(r_+\). These two distinct critical radii organize their respective winding limits.

\section{High-winding path-length limit}
\label{app:high-winding-limit}
\par
Fix the source and observer radii, and let \(S(\Theta)\) be the length of the returning geodesic with positive total angular sweep \(\Theta\). With an arbitrary path parameter \(\lambda\), its arc length is
\begin{equation}
S = \int_{\lambda_i}^{\lambda_o}\mathcal L\,d\lambda, \qquad
\mathcal L = \sqrt{f(r)^{-1}\left(\frac{dr}{d\lambda}\right)^2 + r^2\left(\frac{d\phi}{d\lambda}\right)^2}.
\label{eq:arc-length-functional}
\end{equation}
Along a geodesic, the first endpoint variation of this functional reduces to a boundary term. Because the endpoint radii are fixed, only the total angular sweep is varied. Since \(ds=\mathcal L\,d\lambda\), Eq.~\eqref{eq:conserved-angular-momentum} gives
\begin{equation}
\frac{dS}{d\Theta}=\frac{r^2(d\phi/d\lambda)}{\mathcal L}=r^2\frac{d\phi}{ds}=J(\Theta).
\label{eq:path-length-endpoint-variation}
\end{equation}
The angular coordinate is defined to increase along the chosen propagation orientation, so Eq.~\eqref{eq:path-length-endpoint-variation} applies to the positive anticlockwise and clockwise sweeps used in Eq.~\eqref{eq:multiloop-angles}. The radial turning point is an interior point of the smooth geodesic, rather than an independently varied endpoint, and therefore contributes no boundary term.

\par
For a returning path, \(r_+<J\leq r_{\min}\). If \(J\) remains strictly separated from \(r_+\), the angular accumulation along both radial legs is finite because the square-root singularity at the turning point is integrable. Consequently, an unbounded angular sweep is mathematically possible only when the turning point approaches the throat:
\begin{equation}
\lim_{\Theta\to\infty}J(\Theta)=r_+.
\label{eq:angular-momentum-limit}
\end{equation}

\par
For either winding family, define \(S_n=S(\Theta_n)\). Equation~\eqref{eq:multiloop-angles} dictates \(\Theta_{n+1}-\Theta_n=2\pi\). The mean-value theorem then guarantees a value \(\xi_n\) between \(\Theta_n\) and \(\Theta_{n+1}\) such that
\begin{equation}
\frac{S_{n+1}-S_n}{2\pi}=\left.\frac{dS}{d\Theta}\right|_{\Theta=\xi_n}=J(\xi_n).
\label{eq:mean-value-path-increment}
\end{equation}
As \(n\to\infty\), \(\xi_n\to\infty\). Combining eqs.~\eqref{eq:angular-momentum-limit} and \eqref{eq:mean-value-path-increment} yields
\begin{equation}
\lim_{n\to\infty}(S_{n+1}-S_n) = 2\pi r_+.
\label{eq:appendix-loop-increment}
\end{equation}

\section{Numerical convergence of multi-loop path lengths}
\label{app:path-length-convergence}
\par
For the numerical comparison below, all high-order lengths are extrapolated from the last directly computed path:
\begin{equation}
S_n^{\rm asy}=S_5^{\rm num}+(n-5)2\pi r_+,\qquad n>5.
\label{eq:high-order-asymptotic-length}
\end{equation}

\par
Table~\ref{tab:rn_geodesic_lengths} retains the arbitrary-precision high-order path lengths and reports the relative error of Eq.~\eqref{eq:high-order-asymptotic-length}. The source is fixed at \(r_i=5r_+\), the observer at \(r_o=8r_+\), and both winding orientations are evaluated independently. The reference lengths utilize 60-digit arithmetic, a 96-point Gauss--Legendre rule, and 150 bisection steps in \(y=\ln(J-r_+)\). The largest angular residual is \(5.118\times10^{-44}\,\mathrm{rad}\). An independent 80-point rule changes the displayed \(n>5\) increments by at most \(3.176\times10^{-17}r_g\). For \(n=6,\ldots,9\), the relative error of the high-order approximation does not exceed \(1.377\times10^{-10}\).
\begin{table}[tbp]
\centering
\begingroup
\footnotesize
\renewcommand{\arraystretch}{1.12}
\begin{tabular}{crr@{\quad}crr}
\multicolumn{3}{c}{\(\mathrm{ACW}\)} & \multicolumn{3}{c}{\(\mathrm{CW}\)} \\
$n$ & $S_n^{\rm num}/r_g$ & $\epsilon_n$ & $n$ & $S_n^{\rm num}/r_g$ & $\epsilon_n$ \\
6 & 42.4662046493 & $1.377\times10^{-10}$ & 6 & 44.1417207313 & $3.673\times10^{-11}$ \\
7 & 47.4927528951 & $1.257\times10^{-10}$ & 7 & 49.1682689771 & $3.368\times10^{-11}$ \\
8 & 52.5193011409 & $1.137\times10^{-10}$ & 8 & 54.1948172228 & $3.057\times10^{-11}$ \\
9 & 57.5458493866 & $1.038\times10^{-10}$ & 9 & 59.2213654685 & $2.797\times10^{-11}$ \\
\end{tabular}
\endgroup
\caption{Numerical high-order path lengths and relative error $\epsilon_n=|S_n^{\rm num}-S_n^{\rm asy}|/S_n^{\rm num}$ of the high-order approximation in Eq.~\eqref{eq:high-order-asymptotic-length}. The ACW and CW returning paths are evaluated independently for $r_i=5r_+$ and $r_o=8r_+$.}
\label{tab:rn_geodesic_lengths}
\end{table}

\section{Reference and platform Green's functions}
\label{app:platform-green}
\par
An ideal scalar field confined to the two-dimensional RN spatial slice obeys
\begin{eqnarray}
&&\left(\frac{1}{c^2}\partial_t^2-\Delta_h\right) \Psi(t,\mathbf{x}) = S(t,\mathbf{x}),
\label{eq:surface-wave-equation}\\
&&\Delta_h = \frac{\sqrt{f(r)}}{r}\partial_r \!\left[r\sqrt{f(r)}\,\partial_r\right] +\frac{1}{r^2}\partial_\phi^2 .
\label{eq:surface-laplace-beltrami}
\end{eqnarray}
With the harmonic convention \(e^{-i\omega t}\), the flat two-dimensional outgoing Green function is
\begin{equation}
G_{2,\rm flat}^{(+)}(\omega,L) =\frac{i}{4}H_0^{(1)} \!\left(\frac{\omega L}{c}\right).
\label{eq:flat-2d-outgoing-green}
\end{equation}
In the high frequency regime \(\omega L/c\gg1\), its leading asymptotic form is
\begin{equation}
G_{2,\rm flat}^{(+)}(\omega,L) \simeq\frac{\exp\!\left[ i\left(\omega L/c+\pi/4\right) \right]}{\sqrt{8\pi(\omega/c)L}}.
\label{eq:flat-2d-outgoing-asymptotic}
\end{equation}
The reference kernel used in the main text and the corresponding spreading weights are
\begin{eqnarray}
&&G_0^{\rm ref}(\omega,L_j)= \frac{\exp(i\omega L_j/c)}{L_j},
\label{eq:green-reference-kernel}\\
&&w_j^{(2\rm D)} \propto L_j^{-1/2},
\qquad w_j^{\rm ref}=L_j^{-1}.
\label{eq:green-spreading-weights}
\end{eqnarray}
Thus \(L_j^{-1/2}\) is the flat two-dimensional spreading weight, whereas the main text adopts the three-dimensional point-source weight \(L_j^{-1}\) from Figure~\ref{fig:point-source} applied to the surface-geodesic length as a reference amplitude model~\cite{BornWolf2019,Chew1999}.

\par
For a selected path of length \(L_j\), a platform Green function can be represented schematically as
\begin{equation}
G_{\rm plat}^{(j)}(\omega) =\mathcal{A}_j(\omega) \exp\!\left\{i\beta(\omega)L_j-\alpha(\omega)L_j\right\}.
\label{eq:platform-green-core}
\end{equation}
Here \(\mathcal{A}_j\) accounts for two-dimensional spreading, curvature focusing, caustic phase, source and mode coupling, and boundary normalization; \(\beta\) and \(\alpha\) describe dispersion and attenuation. When the phase of \(\mathcal{A}_j\) varies slowly, the path phase and group delay are given by \(\beta L_j\) and \(\partial_\omega[\beta L_j]\). The RN lengths therefore serve as the geometric backbone, while platform-specific factors govern the measured delays, peak profiles, and fringe contrast~\cite{Plebanski1960,Chew1999,Yariv1973,KristensenQNM2020}.

\acknowledgments
This work is supported by the National Natural Science Foundation of China (Grant No. 12505060), Fapesq-PB of Brazil, the Fund Project of Chongqing Normal University (Grant Number: 24XLB033), Chongqing Natural Science Foundation General Program (Grant No. CSTB2025NSCQ-GPX1019), and Science and Technology Research Project of Chongqing Municipal Education Commission (Grant Number: KJQN202500563).

\FloatBarrier

\bibliographystyle{JHEP}
\bibliography{references}

\end{document}